\documentclass{aa} 
\usepackage{amsmath} 
\usepackage{txfonts}
\usepackage{graphicx} 
\usepackage{url}
\usepackage{natbib} 
\bibpunct{(}{)}{;}{a}{}{,}

\newcommand{\ie}{{\it i.e.}~}
\newcommand{\eg}{{\it e.g.}~}
\newcommand{\Ms}{\ensuremath{M_\odot}}

\newcommand{\el}[2]{$\rm{}^{#2}\kern-0.6pt#1$}
\newcommand{\kms}{km\,s$^{-1}$}

\begin{document}

\title{The evolution of two stellar populations in globular clusters}
\subtitle{I. The dynamical mixing timescale}


\author{T. Decressin\inst{1} \and H. Baumgardt\inst{1} \and P. Kroupa\inst{1}}

\offprints{T. Decressin,\\ email: decressin@astro.uni-bonn.de}

\institute{$^1$ Argelander Institute for Astronomy (AIfA), Auf dem H\"ugel 71, D-53121 Bonn, Germany}

\date{Received / Accepted}

\authorrunning{} \titlerunning{}

\abstract{}%
{We investigate the long-term dynamical evolution of two distinct stellar
  populations of low-mass stars in globular clusters in order to study
  whether the energy equipartition process can explain the high number of
  stars harbouring abundance anomalies seen in globular clusters.}%
{We analyse N-body models by artificially dividing the low-mass stars
  ($m\le0.9$~\Ms{}) into two populations: a small number of stars (second
  generation) consistent with an invariant IMF and with low specific
  energies initially concentrated towards the cluster-centre mimic stars
  with abundance anomalies. These stars form from the slow winds of
  fast-rotating massive stars. The main part of low-mass (first generation)
  stars has the pristine composition of the cluster. We study in detail how
  the two populations evolve under the influence of two-body relaxation and
  the tidal forces due to the host galaxy.}%
{Stars with low specific energy initially concentrated toward the cluster
  centre need about two relaxation times to achieve a complete
  homogenisation throughout the cluster. For realistic globular clusters,
  the number ratio between the two populations increases only by a factor
  2.5 due to the preferential evaporation of the population of outlying
  first generation stars.  We also find that the loss of information on the
  stellar orbital angular momentum occurs on the same timescale as spatial
  homogenisation.}%
{To reproduce the high number of chemically anomalous stars in globular
  clusters by preserving an invariant IMF, more efficient mechanisms such
  as primordial gas expulsion are needed to expel the stars in the outer
  cluster parts on a short timescale.}

\keywords{globular clusters: general - stellar dynamics - Methods: N-body
  simulations}

\maketitle

\section{Introduction}

In a given globular cluster (GC), low-mass stars are known to display
a homogeneous composition in Fe and other heavy elements (with the
noticeable exception of $\omega$~Cen). However, in each individual GC some
pronounced star-to-star variations in light element abundances (Li to Al)
are found, leading to the well-known C-N, O-Na, Li-Na and Mg-Al
anticorrelations \citep[for a review
see][]{GrattonSneden2004,Charbonnel2005}. They are the direct results of
H-burning nucleosynthesis at high temperatures
\citep{DenisenkovDenisenkova1989,DenisenkovDenisenkova1990,LangerHoffman1995,PrantzosCharbonnel2007}.
In addition, since such abundance patterns are not detected in population
II halo fields stars, they are expected to be related to the evolution of
the GCs. Since the low-mass stars still on the main sequence
or on the red giant branch display such abundance variations and since
these stars do not produce in their interior the high temperatures needed
to create the abundance anomalies, it is expected that low-mass stars have
inherited this chemical pattern at birth.

Massive stars as possible sources to contaminate the metal abundances in
the GCs have first been proposed by
\citet{PilachowskiLeep1982} and \citet{WallersteinLeep1987} and then
studied by \citet{PrantzosCharbonnel2006} and \citet{Smith2006}. Recently
\citet{DecressinMeynet2007} explore the role of a high initial rotation in
such stars, finding that abundance anomalies in GCs can
result from an early pollution by fast-rotating massive stars. Indeed
during main sequence evolution, angular momentum is transported from the
centre to the stellar surface, and for stars heavier than 20~\Ms{} with a
high initial rotational velocity, their surface can reach the break-up
velocity limit at the equator (\ie{} the centrifugal equatorial force
balances gravity). In such a situation, a slow mechanical wind develops at
the equator and forms a disc around the stars similar to what happens to Be
stars \citep{TownsendOwocki2004,EkstromMeynet2008}. The second effect of
rotation is to transport elements from the convective H-burning core to the
stellar surface, which enriches the disc with H-burning material. While
matter in discs has a slow outward velocity and hence will stay in the
potential well of the cluster, matter released during the main part of the
He-burning phase and during SN explosions has a very high radial velocity
and will be lost from the cluster. Therefore, new stars can only form from
the matter available in discs and can become the stars with abundance
anomalies we observe today. Thus GCs can contain two distinct
populations of low-mass stars: a first generation that has the chemical
composition of the material out of which the cluster formed (similar to
field stars with a similar metallicity); and a second generation of stars
harbouring the abundance anomalies born from the ejecta of fast-rotating
massive stars.

Based on the determination of the composition of giant stars in NGC~6752
by \citet{CarrettaBragaglia2007}, \citet{DecressinCharbnnel2007} determined
that around 85\% of stars (of the sample of 120 stars) display abundance
anomalies. \citet{PrantzosCharbonnel2006} find similar results for NGC~2808
with their analysis of the data of \citet{CarrettaBragaglia2006}: 70\% of
the cluster stars present abundance anomalies. 

A main problem is that by assuming a \citet{Salpeter1955} IMF, the accumulated
mass of the slow winds ejected by the fast-rotating massive stars would
only represent 2.5\% of the mass of the whole population of first
generation stars (see Fig.~5 in \citealp{DecressinCharbnnel2007}). This
number is only a lower limit since some mixing between the slow wind and
the gas of the interstellar medium left after the formation of first
generation stars needs to be taken into account so that more matter will be
available to form second generation stars. Taking this dilution process
into account, \citet{DecressinCharbnnel2007} found that the initial mass of
the second generation stars is 5\% of the cluster mass. As the whole
population of low-mass stars represents half of the cluster stellar mass,
this leads to a number ratio of about 0.1 between the second and the first
generation of low-mass stars. This dilution process of the slow wind in the
ISM is also mandatory to explain the Li detected in stars with abundance
anomalies. Indeed in NGC~6752, \citet{PasquiniBonifacio2005} found that
turn-off stars with abundance anomalies display a noticeable abundance
variation in Li. Similar results are found for 47~Tuc
\citep{BonifacioPasquini2007}. However in the hot interior of massive
stars, all the Li initially present is consumed and no Li is ejected into
the disc. Thus the Li in stars with abundance anomalies can only come from
the ISM.

To reconcile the high fraction of second generation stars with the
observations \citet{DecressinCharbnnel2007} suggest that we need either (a)
a flat IMF for the massive stars, or (b) that 96\% of the low-mass stars of
the first generation have escaped the cluster. The second case could be
achieved if the following scheme happens: we assume that the fast-rotating
massive stars are born near the cluster centre or migrate rapidly toward
the centre (\ie, that the cluster is initially mass segregated). In such a
case, the slow winds ejected by these massive stars are also concentrated
near the centre. As the second generation stars need to be formed on a
short timescale to avoid the ensuring supernovae pushing the slow wind
material, from which second generation stars are to form, out of the
cluster, we can expect that second generation stars will form with the same
radial distribution as the massive stars.

Primordial mass segregation is observed in young star clusters in the Milky
Way and the Magellanic Clouds
\citep{BonnellDavies1998,GouliermisKeller2004,ChendeGrijs2007}. Thus we
could reasonably expect that GCs were also mass segregated at
birth. It should be noted that the recent correlation between the slope of
the mass function and the concentration parameter ($c = \log (r_t/r_c)$,
where $r_t$ and $r_c$ are the tidal and core radius respectively)
discovered by \citet{DeMarchiParesce2007}, can be explained naturally if
the GCs started mass segregated
\citep{MarksKroupa2008,BaumgardtDeMarchi2008}.

Empirical evidence for the presence of gas in the cluster centre at the
same time as massive stars are still evolving is provided by
\citet{GallianoAlloin2008}. They investigated extremely massive young
clusters, which are believed to be nascent GCs (around 6-8
Myr old), in the spiral Galaxy NGC 1365, and found an optically thick
component in addition to a thin one. These authors argue that this thick
component could be related to a subsequent or on-going episode of star
formation.

We expect that the slow winds of fast-rotating massive stars dilute the ISM
locally around each massive star, so second generation stars will be born
only in the central part of the cluster while the outer regions are
populated exclusively by first generation stars. If some dynamical process
unbinds the external parts most of the first generation stars may be lost
while the second generation stars are retained in the cluster. In the
present paper we investigate if the tidal interactions with the host Galaxy
can produce such an effect.

The time needed for the second generation of stars to spread throughout the
cluster is expected to scale with the two-body relaxation time
at the half-mass radius, $t_\text{rh}$, of the cluster given by
\citet{Spitzer1987},
\begin{equation}
  \label{eq:trel}
  \frac{t_\text{rh}}{1\ \text{Myr}} = \frac{664}{\ln\left(0.02 N\right)}
  \sqrt{\frac{M_\text{Cl}}{10^5\ \Ms}} \left(\frac{\langle m\rangle}{1\
      \Ms}\right) \left(\frac{r_\text{h}}{1\ \text{pc}}\right)^{1.5},
\end{equation}
where $N$, and $\langle m\rangle$ are the number of stars and their mean
mass while $M_\text{Cl}$, and $r_\text{h}$ represent the total mass and the
half-mass radius of the cluster respectively. The two-body relaxation time
is an estimate of the timescale over which the cluster rearranges itself
significantly due to its evolution towards energy equipartition as driven
by the many encounters between two stars within the cluster potential.  At
the same time, the tidal forces of the host galaxy remove stars from the
outer cluster parts. Current observations of GCs suggest that
most of them are several relaxation times old, so that we would expect that
stars in these clusters are thoroughly mixed and therefore
homogeneous at the present time.

In this paper, we study more quantitatively the interplay between
the effect of internal dynamics and the external potential of the host
galaxy on the evolution of two stellar populations in GCs. In
\S~2, we describe the dynamical models we used. Then our analysis is
presented in \S~3. Finally, our results are discussed and our conclusions
are drawn in \S~4. 

\section{Dynamical models}

\subsection{Used models}

The analysis presented here is based on the calculations of
\citet{BaumgardtMakino2003}, which consist of a set of models containing 
 8k to 128k stars\footnote{In this paper we
  use the computer unit 1k = 1024.} initially, carried out with the collisional
Aarseth N-body code \textsc{nbody4} \citep{Aarseth1999}. It uses a
Hermite scheme with individual time-steps for the integration and treats
close encounters between stars by KS \citep{KustaanheimoStiefel1965} and
chain regularization \citep{MikkolaAarseth1990,MikkolaAarseth1993}.

Here we restrain the analysis to clusters moving on circular orbits (with
orbits 8500~pc away from the Galactic centre) through a host galaxy
modelled as a logarithmic potential. The clusters follow King profiles
initially with central concentrations $W_0=5.0$. The cluster initial radii
are adjusted such that the tidal radius of the King model was equal to the
tidal radius given by the host Galactic tidal field. The clusters are
computed until complete dissolution. The initial mass function of stars in
the cluster is given by a canonical two-part power-law mass function
\citep{Kroupa2001} with upper and lower mass limits of 15 and 0.1~\Ms{}
respectively. This leads to an initial mean mass of $\langle m \rangle =
0.547$~\Ms{}. Stellar evolution is modelled by the fitting formulae of
\citet{HurleyPols2000} with an assumed metallicity of $Z=0.001$
($\text{[Fe/H]}\simeq -1.2$, the mean value of GC
metallicities) for the cluster stars. Mass lost from the stars in the
course of their evolution is assumed to be immediately lost by the
cluster. The clusters contained no initial binaries.

\subsection{Selection by specific energy}

Although the models of \citet{BaumgardtMakino2003} have been computed for a
single stellar population, we apply a process which could mimic the
formation of a cluster with two dynamically distinct
populations. \citet{DecressinCharbnnel2007} have shown that the slow
equatorial winds of fast-rotating massive stars create slowly expanding
discs which stay in the potential well of the GCs.  If we assume a
radial velocity for these winds between 1 and 10~\kms, and that the gas
travels for about $10^5$~yr before starting to form new stars, it will
travel for a distance of about 0.1 to 1~pc. As the initial tidal radius of
the cluster is a few 10~pc, we expect that second-generation stars have the
same initial spatial distribution as their massive progenitors. Since we
assume here that the massive stars are born in the centre of the cluster,
the second generation stars will also be located near its centre.

In order to select stars in the cluster centre we sort all the low-mass
stars ($M\le 0.9$~\Ms) according to their specific energy defined as the
energy per unit mass. We define the second stellar generation as the stars
with smallest specific energies, (\ie{} those stars which are most tightly
bound to the cluster due to their small central distance and low
velocity). The number of second generation stars is given by having their
total mass represent 5\% of the initial mass of the cluster. For the models
with initially 128k stars, we have 104691 first generation stars and 10327
second generation stars; the remaining stars are more massive than
0.9~\Ms{}.

\subsection{Possible issues}

\begin{figure}
  \includegraphics[width=.5\textwidth]{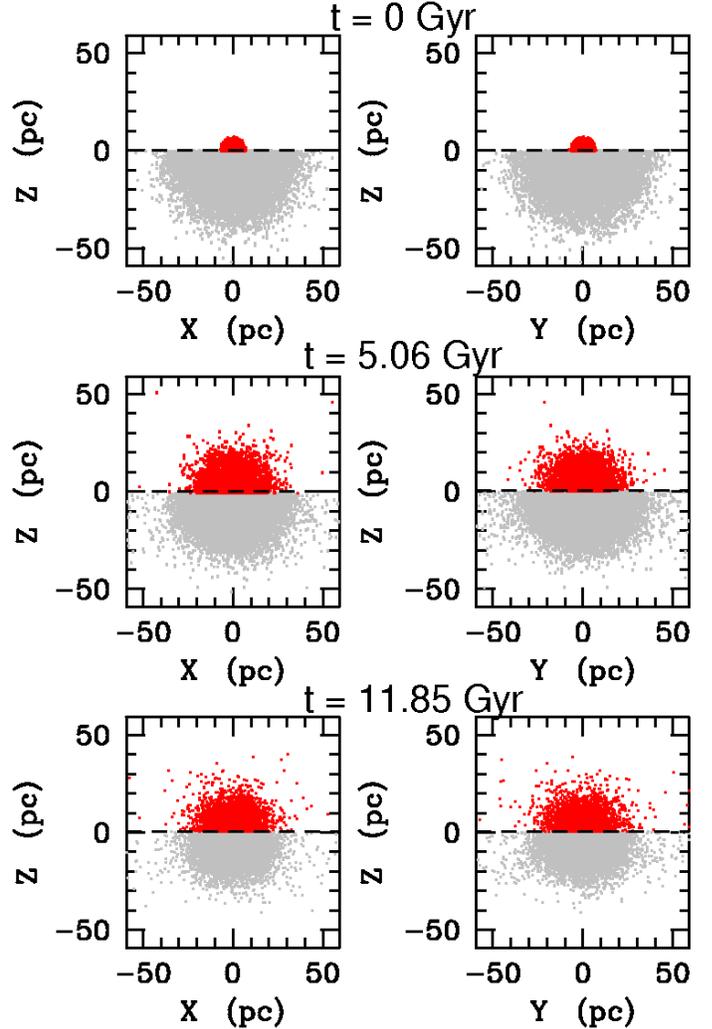}
  \caption{Position of low-mass stars ($M_\text{ini}\le0.9$~\Ms) in the
    Z-X (left) and Z-Y (right) plane at different times: 0~Gyr (top),
    5.06~Gyr (middle) and 11.85~Gyr (bottom) for the N-body integration with
    initially 128k stars. For sake of clarity, the upper and bottom parts
    of each panel display only the second and first generation of stars
    respectively.}
  \label{fig:posi}
\end{figure}

\begin{figure}
  \includegraphics[width=.5\textwidth]{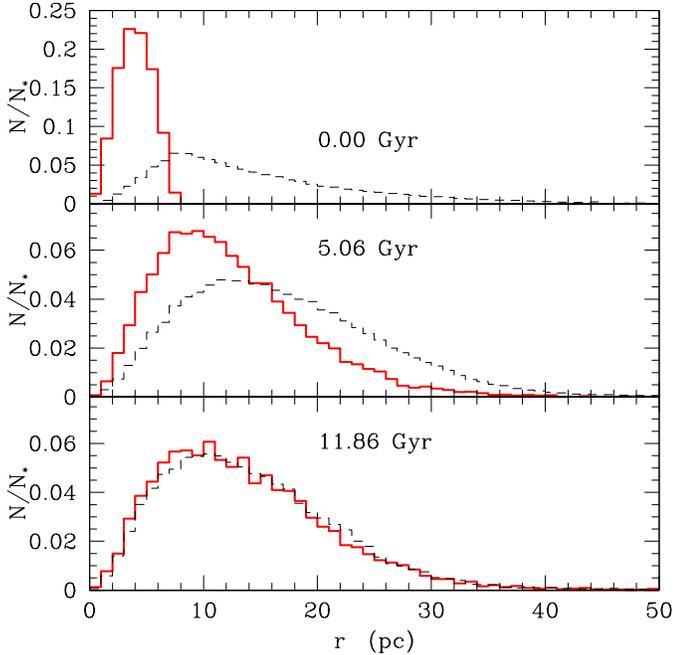}
  \caption{Radial distribution of the first (dashed lines) and second
    (full lines) generation of low-mass
    stars at the same times as in Fig.~\ref{fig:posi}. Each histogram is
    normalised to the total number of stars in each population.}
  \label{fig:distrr}
\end{figure}

It should be noted that the upper-limit for the IMF in the models of
\citet{BaumgardtMakino2003} is 15~\Ms{}. However the pollution by massive
stars would involve stars more massive than 20~\Ms{}
\citep[see][]{DecressinMeynet2007}.  The evolution of such high mass stars
takes a few million years. So if the delay for the formation of the second
generation stars is short, the two populations will be nearly
coeval. However the direct dynamical effects of the fast-rotating massive
stars (more massive than 20~\Ms{}), like the lowering of the cluster
potential well by the ejection of their fast polar winds and SN, or the
creation of a second generation of stars with the matter released in their
slow winds, are not taken into account in this paper.

With the scheme presented in \S~2.2, we implicitly assume that the lifetime
of the first and second generation of low-mass stars are identical. In real
clusters this would not be valid if a noticeable fraction of second
generation stars shows a large overabundance in He. Indeed as He is the
main product of H-burning in stars, abundance anomalies are also expected
to come with He overabundance. Although He abundance cannot be directly
determined, several indirect observations push toward this finding. First,
the GCs $\omega$~Cen and NGC~2808 display multiple main
sequences \citep{PiottoVillanova2005,PiottoBedin2007}; a double sub-giant
branch is also found in NGC~1851 \citep{MiloneBedin2008}. Such features can
only be understood if the stars present various He contents. He enrichment
is also a possible explanation for the appearance of extreme horizontal
branches seen for several GCs
\citep{CaloiD'Antona2005,CaloiD'Antona2006}.

For our paper, the different initial He content between the two populations
needs to be properly taken into account as stars with higher initial He
content have a shorter lifetime due to faster stellar evolution. Therefore
to be consistent with the evolution of the second generation of stars
requires custom-made models. However the value of $Y=0.4$ (instead of 0.25)
is often cited as needed to properly explain the observations
\citep{Norris2004,PiottoVillanova2005} while \citet{DecressinCharbnnel2007}
find that fast-rotating massive stars could even create low-mass stars with
a He mass fraction up to $Y = 0.7$. For 12~Gyr old clusters,
\citet{DAntonaVentura2007} show that the turn-off mass decreases from 0.8
to 0.64~\Ms{} when the initial He content raises from 0.24 to 0.4. It could
even be as low as 0.35~\Ms{} if the initial He content is 0.7 (Decressin et
al., in prep.). Unfortunately the N-body computations of
\citet{BaumgardtMakino2003} use only standard He abundance for the stellar
evolution models, so this effect cannot be investigated in the present
paper. However we expect that the global findings of this paper on the
dynamical timescales will not be changed by this He-rich
population. Nevertheless a forthcoming paper will address this issue.

\section{Analysis}

\begin{figure}
  \includegraphics[width=0.5\textwidth]{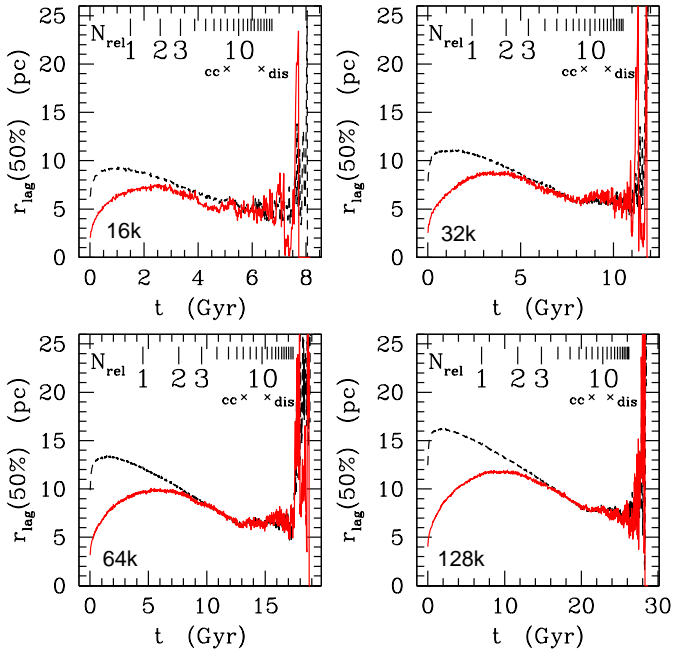}
  \caption{Evolution of the half-mass radius for first (dashed lines) and
    second (full lines) generation of low-mass stars present in the
    clusters as a function of time for the models with initially 16k (top
    left), 32k (top right), 64k (bottom left) and 128k (bottom right)
    stars. At the top of each panel the number of passed two-body relaxation times is
    shown (Eq.~(\ref{eq:Nrel})), and crosses indicate the time of core
      collapse and of cluster
  dissolution.}
  \label{fig:rmean}
\end{figure}

\subsection{Dynamical evolution: radial distribution}

Fig.~\ref{fig:posi} illustrates the evolution of two selected stellar
populations in a cluster with initially 128k particles in a circular orbit
around the Galaxy at three different times. In Fig.~\ref{fig:distrr} one
can see the radial distribution of the two populations at the same
epochs. Initially, the second generation stars with low specific energy are
concentrated within 6~pc around the centre of the cluster while stars of
the first generation show a more extended distribution up to
40~pc.

Progressively the second generation stars spread out due to dynamical
encounters so their radial distribution extends. The middle panels in
Fig.~\ref{fig:posi} and \ref{fig:distrr} show that even after 5~Gyr of
evolution the two populations still retain different distributions. At this
time, which corresponds roughly to one initial relaxation time for the whole
cluster, the extent of the second generation stars reaches 35~pc, but
they are still more strongly concentrated to the central part.

The bottom panels in Fig.~\ref{fig:posi} and \ref{fig:distrr} show that after
nearly 12~Gyr of evolution (slightly more than 2 initial relaxation times) the
two populations have similar radial distributions and can no longer be
distinguished on the basis of their dynamical properties.  

\subsection{Number of escaping stars}

As already stated in the introduction, two competitive processes act in
the clusters: the loss of stars from the outer cluster parts will first
reduce the number of bound first generation stars; and the dynamical spread
of the initially more concentrated second generation stars will stop this
differential loss when the two populations are dynamically mixed.  We now
describe the operation of this interplay.

The time evolution of the half-mass radii of the two populations is
displayed in Fig.~\ref{fig:rmean} for clusters with different initial
numbers of particles. Stars which have already been lost by the clusters
are not taken into account, so this radius reflects only bound
stars. Initially, stars of the first generation have a larger half-mass
radius as expected by our selection scheme (\eg{} 5 and 12~pc respectively
for the case with $N = 128$k stars). The fast initial increase of the
half-mass radius is due to the evolution of the massive stars which lose a
large part of their mass (Fig.~1 in \citealp{BaumgardtMakino2003}). Winds
from these stars are not retained so the potential well of the cluster is
lowered and the outer regions expand. Later the tidal forces of the host
galaxy remove outer stars and the half-mass radius decreases until the
final disruption of the cluster. The strong variation at the end of the
evolution is due to the small number of low-mass stars still remaining in
the cluster. They are also more pronounced in the smaller clusters.

On the other hand, the second generation stars have initially a smaller
half-mass radius as a result of their lower specific energy. The initial
expansion due to the evolution of massive stars has a smaller effect on the
half-mass radius of the second generation stars compared to first
generation stars as second generation stars populate the centre and hence
the deeper and less perturbed parts of the potential well. Later the
half-mass radius increases continuously with time following the spread of
the radial distribution as the cluster tries to achieve energy
equipartition. After some time the two radii are similar and the further
evolution of the two populations is identical.

Fig.~\ref{fig:rmean} also shows the number of passed two-body relaxation times,
$N_\text{rel}$, defined as:
\begin{equation}
  \label{eq:Nrel}
  N_\text{rel}(t) = \int_0^t\frac{d\tilde t}{t_\text{rh}(\tilde t\,)},
\end{equation}
where $t_\text{rh}(\tilde t)$ is the current relaxation time given by
Eq.~(\ref{eq:trel}) and $\tilde t$ is the integration variable. The
relaxation time, $t_\text{rh}$, decreases with time as both $r_\text{hm}$
and N decrease (see Eq.~(\ref{eq:trel})).  For each cluster we also
indicate the time at which core-collapse occurs and the time at which
dissolution of the cluster happens (\ie{} when 95\% of the initial mass has
been lost by the cluster) as computed by \citet{BaumgardtMakino2003}. These
two times are labelled by ``cc'' and ``dis'', respectively. Cluster
homogenisation needs around 2 relaxation times nearly independently of the
initial number of stars. However there is still a weak dependence with the
initial number of stars since the effect of mass loss does not scale with N
in the same way as the relaxation time \citep{Baumgardt2001}. Other radial
properties, like the mean radius and the radial distribution of stars, are
also similar in both populations when $N_\text{rel} \gtrsim 2$. Since
GCs are several relaxation times old, they have lost all
information on their initial radial distribution. The case of $\omega$~Cen,
which is dynamically younger, will be discussed in \S~3.4

\begin{figure}
  \includegraphics[width=0.5\textwidth]{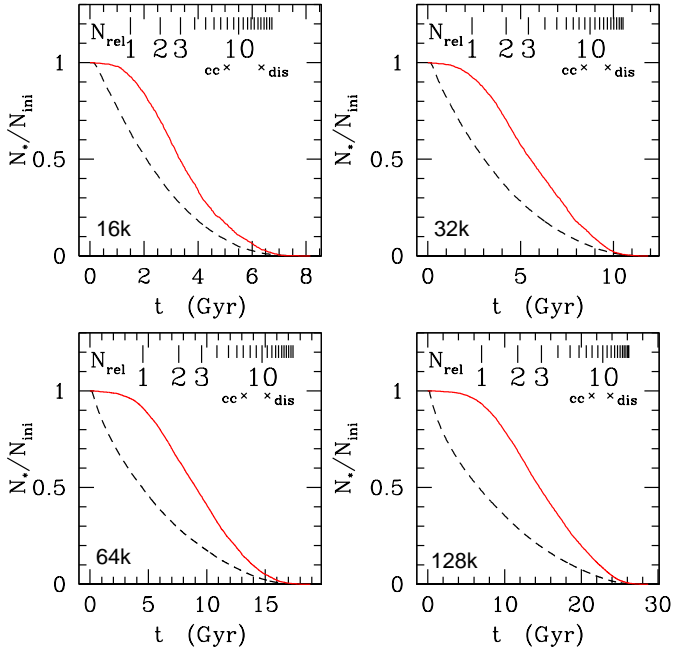}
  \caption{Evolution with time of the ratio of first generation (dashed
    lines) and second generation (solid lines) low-mass stars to their
    initial content for the models with 8k (top left), 32k (top right), 64k
    (bottom left) and 128k particles (bottom right). At the top of each
    panel the number of passed two-body relaxation times is shown
    (Eq.~(\ref{eq:Nrel})), and crosses indicate the time of core collapse and of
  cluster dissolution.}
  \label{fig:mloss}
\end{figure}

\begin{figure}
  \includegraphics[width=0.5\textwidth]{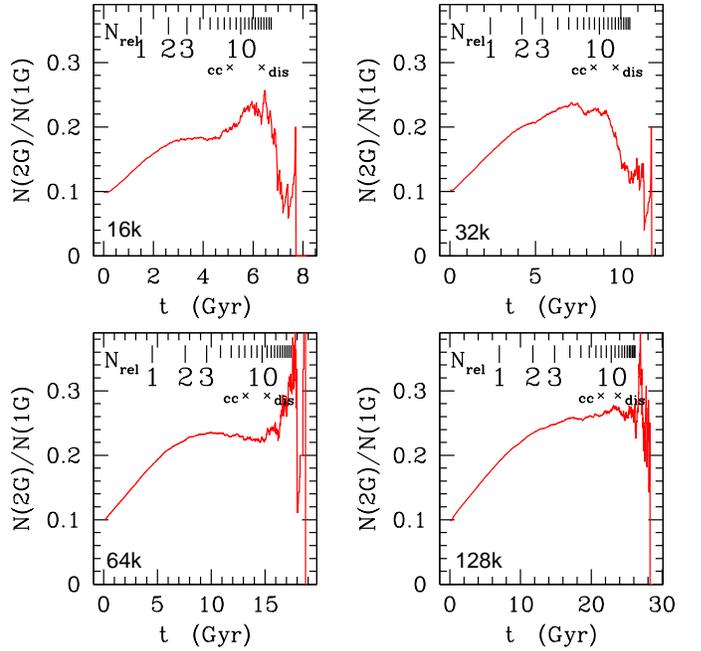}
  \caption{Number ratio between the second (with low initial specific
    energy) and first (with high initial specific energy) population of
    low-mass stars in a cluster as a function of time for the same
    models as in Fig.~\ref{fig:mloss}. At the top of each panel the
    number of passed two-body relaxation times is shown (Eq.~(\ref{eq:Nrel})),
    and crosses indicate the time of core collapse and of
  cluster dissolution.}
  \label{fig:mratio}
\end{figure}

As previously seen, the effect of the external potential of the Galaxy on
the cluster is to strip away stars with high specific
energy. Fig.~\ref{fig:mloss} shows the evolution of the fraction of stars
in each population that still remains in the cluster as a function of time
for different initial numbers of stars.  Initially, only stars of the first
generation populate the outer part of the cluster owing to their high
specific energy. Therefore only first generation stars are lost in the
beginning. This lasts until the second generation stars migrate into the
outer part of the cluster. Depending on the cluster mass, it takes between
1 to 4~Gyr to start a loss of second generation stars. This timescale
scales roughly with the relaxation time (see also \S~3.3 for more details).
Indeed, as shown before, more massive clusters have a more extended initial
radial distribution so the migration of the second generation stars lasts
longer. Besides, these clusters also have a deeper potential well and the
erosion by the tidal forces of the Galaxy is reduced.

Due to the time-delay to lose second generation stars, their remaining
fraction in the cluster is always higher than that of the first
generation stars except during the final stage of cluster dissolution.
Fig.~\ref{fig:mratio} quantifies this point by showing the time evolution
of the number ratio of second to first generation stars. As a direct
consequence of our selection procedure, the initial ratio is about 0.1; and
it then increases gradually with time. It tends to stay nearly constant as
soon as the two distributions are similar. Finally, at the time of cluster
dissolution, large variations occur due to the low number of low-mass stars
present in the cluster.

Over the cluster history, the fraction of second generation stars relative
to first generation ones increases by a factor of 2.5. Therefore, when the
two populations have the same radial distribution, these stars represent
25\% of the low-mass stars present in the clusters. Presently, observed
ratios have been derived only spectroscopically for NGC~2808 (70\%) and
NGC~6752 (85\%), since a large number of stars needs to be analysed. Thus,
the internal dynamical evolution and the dissolution due to the tidal
forces of the host Galaxy are not efficient enough to produce the high
observed value. An additional mechanism is thus needed to expel the first
generation stars more effectively.

\subsection{Influence of the Initial number ratio}

\begin{figure}[t]
  \includegraphics[width=.5\textwidth]{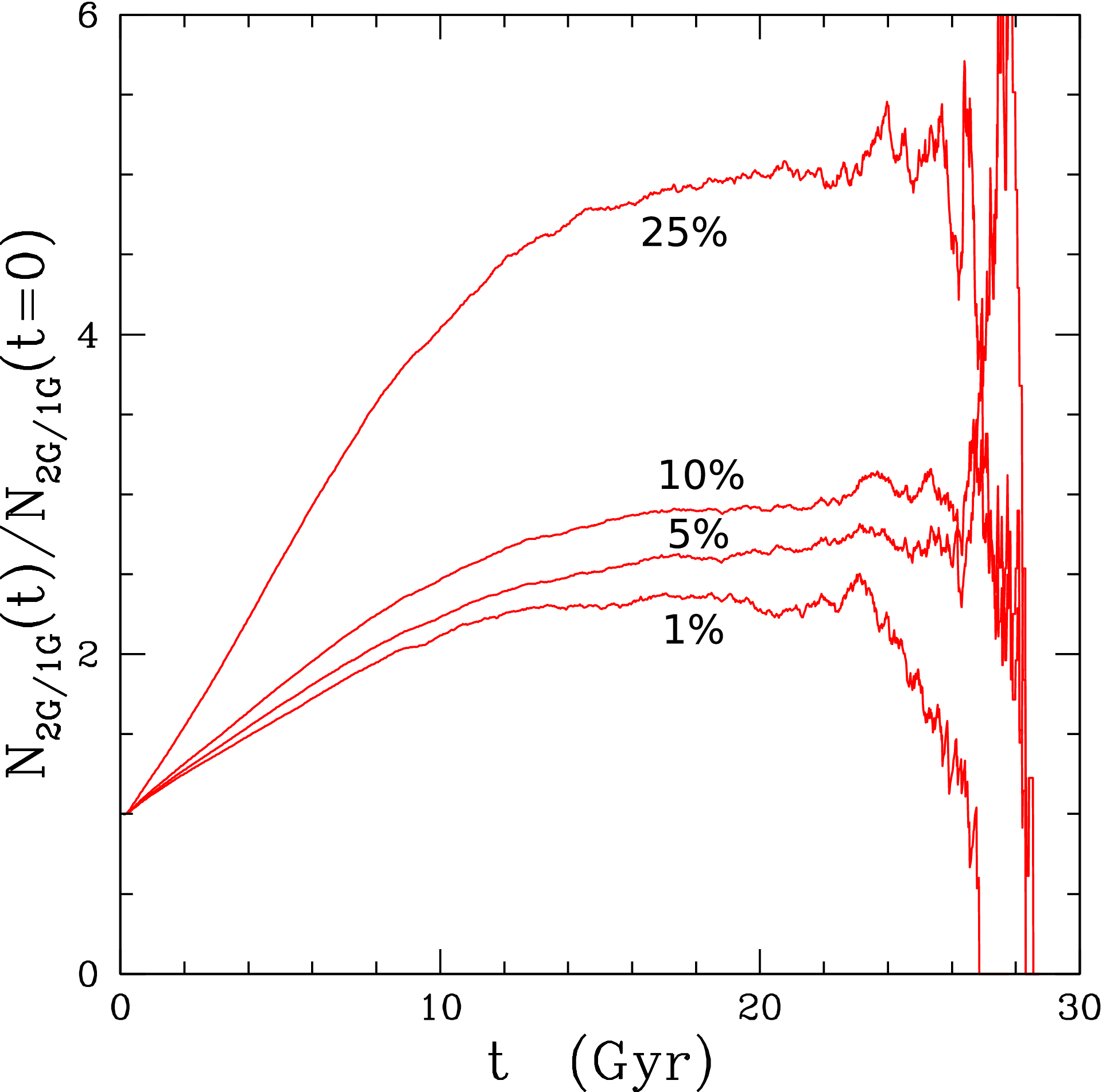}
  \caption{Increase of the ratio of second to first generation stars as a
    function of time for different numbers of initially selected stars 
    for the model with $N = 128$k stars. 1, 5, 10, 25\% refers to
    fraction of the cluster mass populated with second generation stars.}
  \label{fig:varinit}
\end{figure}

The next point we want to investigate is how our results depend on the
number of stars selected to be first or second generation
stars. Fig.~\ref{fig:varinit} addresses this issue: for the computation
with initially 128k stars, we select different number of second-generation
stars which represent 1, 5, 10, and 25\% of the initial mass of the cluster
and follow how the ratio of second to first generation of stars evolves
with time. For the 5\% case, we retrieve our previous results: the ratio
increases by a factor of 2.5 and then flattens until cluster
dissolution. The more numerous the second generation stars are, the
stronger the increase is. In the case of 25\% of stars selected, the ratio
increases by a factor of 5. This trend can be understood as a result of the
selection process which selects mainly those stars which remain bound to
the cluster until the end. However this case can only be achieved if the
IMF slope for the massive stars is around 1.7 (when Salpeter value is
2.35).

For a standard IMF we expect only a limited number of low-mass stars to be
second generation stars. Thus the models above with a low fraction of
second generation stars are more realistic to what
happens in real GCs and all of them predict only a modest
increase of the fraction of the second generation stars.

\subsection{Angular-momentum-selected sub-populations} 

\begin{figure}[ht]
  \includegraphics[width=0.5\textwidth]{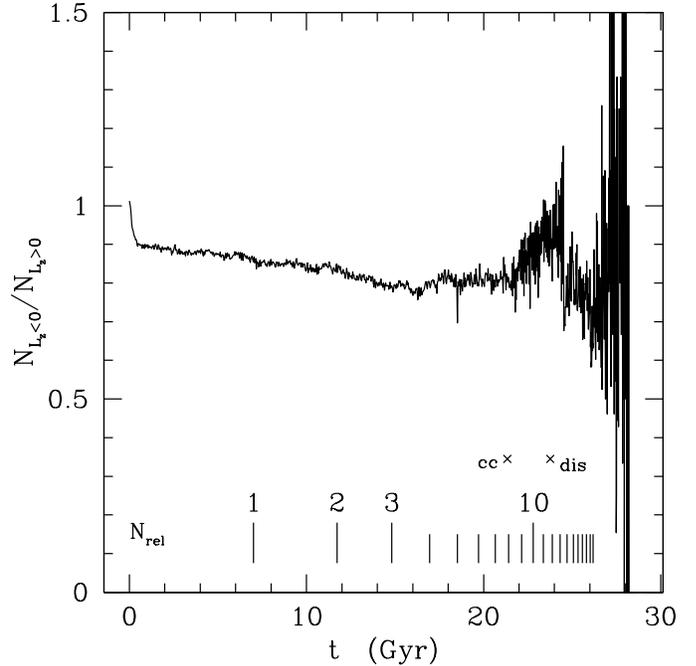}
  \caption{Ratio of stars with negative
    to positive orbital angular momentum in the cluster as a function of
    time. At
    the bottom, the currently passed relaxation time is shown
    (Eq.~(\ref{eq:Nrel})), and crosses indicate the time of core collapse and of
  cluster dissolution.}
  \label{fig:Lzpop}
\end{figure}

\begin{figure}[ht]
  \includegraphics[width=0.5\textwidth]{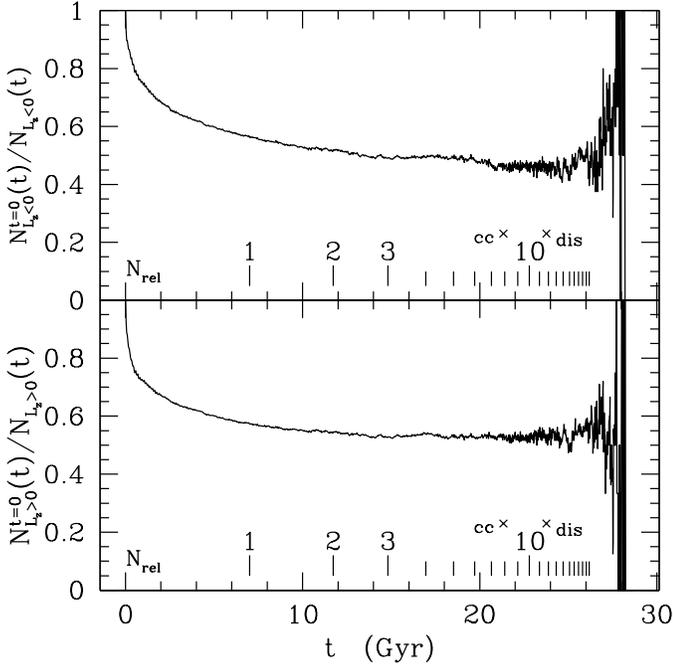}
  \caption{Fraction of stars which have kept their initial orbital
      angular momentum for negative (top) and positive (bottom) cases as a
      function of time. At
    the bottom, the currently passed relaxation time is shown
    (Eq.~(\ref{eq:Nrel})), and crosses indicate the time of core collapse and of
  cluster dissolution.}
  \label{fig:Lzpopin}
\end{figure}

In the previous sections we have only considered the case of two
populations which initially differ by their spatial distribution as we
expect such differences if a second generation of stars are born from the
ejecta of massive stars. In this context, the GC $\omega$~Cen
is an exceptional case as it not only shows variations in light elements
but also a wide spread in metallicity from $\text{[Fe/H]} = -2$ to $-0.5$
\citep{NorrisDaCosta1995}. Additionally, a close inspection of the
color-magnitude diagram of $\omega$~Cen shows that several populations
exist at all the evolutionary stages, from the RGB
\citep{LeeJoo1999,PancinoFerraro2000,SollimaFerraro2005a} to the sub-giant
\citep{FerraroSollima2004,SollimaPancino2005b,VillanovaPiotto2007} branches
and main sequences \citep{BedinPiotto2004,PiottoVillanova2005}.
Interestingly, \citet{NorrisFreeman1996} found the more metal-rich stars to
be more spatially concentrated. Evidence for significant differences in the
kinematics and the spatial distribution between the metal-rich and other
stars have been collected \citep{NorrisFreeman1997,Jurcsik1998}. However
these findings have been challenged by \citet{PancinoGalfo2007} on the
basis of an analysis of a larger sample. 

Therefore in this section we want to determine if the time needed for a
cluster to lose its initial kinematic properties is the same as for the
initial radial distribution. In particular we want to test how quickly the
kinematic differences are erased since differences are observed in
$\omega$~Cen whereas other GCs do not show such strong
observational evidences.  To assert this timescale for achieving kinematic
homogenisation, we setup the initial conditions for the analysis in another
way: we divide the low-mass stars according to their angular momentum about
the cluster centre along the z-axis (the cluster orbital motion occurs in
the x-y plane with positive angular momentum).

First, Fig.~\ref{fig:Lzpop} shows the evolution of the number ratio
between stars having negative and positive orbital angular
momentum. Initially half of the low-mass stars populate the groups with
prograde ($L_z > 0$) and retrograde ($L_z < 0$) motion. A rapid drop of the
ratio of the number of stars with retrograde motion (from 1 to 0.95) occurs
due to the preferential loss of stars with negative $L_z$ during the
evolution of the massive stars, followed by a very slow decline during the
cluster evolution. At the time of core collapse, this ratio is around 0.9.
This fast initial change can be explained by the mass-loss of massive stars
which lowers the potential well of the cluster leading to the expulsion of
the stars in the outer parts as we have seen before. This ejection is more
pronounced for the stars with retrograde motion ($L_z < 0$) relative to the
cluster orbit than those with prograde motion ($L_z > 0$).  This finding
seems to contradict previous works which found that retrograde orbits are
more stable than prograde ones if the cluster is on a prograde orbit in a
steady external potential \citep[see e.g.][]{FukushigeHeggie2000}.  The
difference can be explained by the different coordinate systems used in
these studies: \citet{BaumgardtMakino2003} applied an accelerated and
non-rotating coordinate system (the Galaxy rotates about the cluster) while
an accelerated and rotating one is used by the computations of \citet[in
the cluster centric frame the Galaxy is fixed; the cluster-centric frame is
therefore non-inertial]{FukushigeHeggie2000}. Retrograde orbits in a
rotating frame \citep{FukushigeHeggie2000} are found to be the stable ones,
and a few of them shift to prograde orbits if we consider them in a
non-rotating frame.  This effect leads to the inversion of the dominant
population and explains the discrepancy of \citet{BaumgardtMakino2003} with
the older study. We conclude that stars on retrograde orbits with respect
to their cluster centre are preferably lost from the cluster, where
prograde orbits are in the same sense as the cluster orbit about the
Galaxy. So an initially non-rotating cluster begins to rotate in a prograde
manner after the massive stars have evolved. But the effect remains small
(see Fig.~\ref{fig:Lzpop}). Similar results have been obtained by
  \citet{YimLee2002} who state that clusters with no initial rotation can
  gain a slight positive angular momentum (\ie{} that the cluster rotates
  in the same direction that the orbital motion.).

To measure the time needed to lose the initial angular momentum
distribution, we need to look more carefully at each population, as due to
stellar encounters stars can change their orbital parameters and hence
invert their orbital angular momentum. Initially only two populations
exist: stars with $L_z(t=0)<0$ and $L_z(t=0)\ge0$. As time proceeds,
stellar interactions induce changes to the orbits of stars and some stars
show an inversion in $L_z$. Hence the cluster splits into four populations
depending on their initial and current angular momentum.
Fig.~\ref{fig:Lzpopin} shows the evolution of the fraction of stars which
have kept their initial orbital angular momentum for the population of
stars with negative and positive $L_z$. Here homogenisation will be
obtained when 50\% of the stars have inverted their orbital angular
momentum.  Fig.~\ref{fig:Lzpopin} displays the fraction of stars keeping
their initial AM sign rapidly decreases to 70-80\%, thus 20 to 30\%
of stars suffer such an inversion during the early radial expansion
following the evolution of the massive stars. Then the inversion rate
decreases slowly and homogenisation fully occurs after around 2-3
relaxation times. Thus the loss of information about orbital angular
momentum follows the same timescale as the radial distribution.

Therefore if there is any dynamical (radial or kinematic) property which
differs initially between different cluster sub-populations, most clusters
would have erased them during their evolution as they are several
relaxation times old.  $\omega$~Cen has the longest relaxation time which
spans a range from several $10^9$~yr at the centre to $10^{10}$~yr at the
half-mass radius \citep{vandeVenvandenBosch2006}. Thus it has evolved only
for about one relaxation time and we should indeed see differences, at
least at a low level, if the cluster was born with dynamically distinct
populations. Thus if the results of \citet{PancinoGalfo2007} are confirmed
we would have to conclude that the different populations observed in the
CMD were born with the same radial and kinematic properties.

\subsection{Timescale for early processes}
 
\begin{figure}
  \includegraphics[width=.5\textwidth]{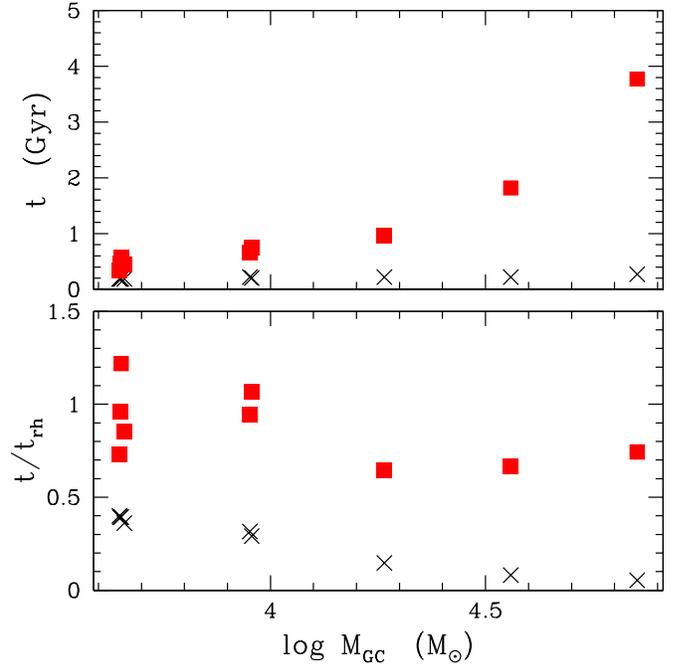}
  \caption{(Top) Time needed for 1\% of the first generation (crosses) and
    second generation (full squares) of low-mass stars to be lost by the
    cluster as a function of the initial mass of the cluster. (Bottom)
    Ratio of these timescales to the initial relaxation time.}
  \label{fig:time}
\end{figure}

Fig.~\ref{fig:time} gives the timescales for a cluster to lose 1\% of both
generations of low-mass stars as a function of the mass of the
cluster. This time corresponds to the end of the initial plateau seen in
Fig.~\ref{fig:mloss} for the second generation. Since
\citet{BaumgardtMakino2003} computed several N-body models with 8k and 16k
stars, we have included all the data points.

In these models, the first generation stars are the first to be noticeably
lost after around 100~Myr and this is nearly independent of the cluster
mass. On the other hand, the time until 1\% of the second generation
evaporates increases with the cluster mass because these stars have to
travel to a larger radius before being lost. For the heaviest clusters this
happens only after 4~Gyr (this time scales as 0.7--0.8 times the relaxation
time, see bottom panel of Fig.~\ref{fig:time}). It should be noted that the
lightest clusters show a strong statistical effect due to the small number
of second generation stars initially present. As heavier clusters show an
almost identical time relative to their relaxation timescale, we can quite
confidently extrapolate our results to more massive clusters.

As already explained, we need to find an additional mechanism which would
more vigorously eject first generation stars and at the same time keep the
second generation stars. Several processes can play this role. First, disk
or bulge shocking has been argued to heat up the outer part of the cluster
when it passes through the Galactic plane.  This effect increases the loss
of stars.  As the timescale between two passages in the Galactic plane is
around $10^8$~yr and the appearance of second generation stars in the outer
regions needs a few $10^9$~yr, a high number of shocks affecting nearly
exclusively the first generation should have happened. Such process can
also lead to the destruction of many GCs over the Galactic history
\citep{ShinKim2008}.  However old GCs formed about 12~Gyr ago whereas the
Galactic disc formed continuously from 10~Gyr ago to the present. As we
need an efficient physical mechanism which can act at the early time of GC
evolution, disc shocking can be discarded to remove preferentially first
generation stars.

Initial gas expulsion by massive stars
operates early in the cluster history (a Myr after cluster formation
or earlier, \citealp{BaumgardtKroupa2008}). As the residual gas still
present after star formation is removed, the potential well of the
cluster can be strongly reduced and the outer parts of the cluster can
become unbound. This process has already been used successfully by
\citet{MarksKroupa2008} to explain the challenging correlation between the
central concentration and the mass function of GCs as found
by \citet{DeMarchiParesce2007}.

Gas expulsion is expected to proceed on a short timescale similar to the
evolution-timescale of massive stars, and will take place well before any
changes due to the internal dynamics or the tidal interaction with the host
Galaxy, the two processes investigated in this paper. However the results
presented in the previous sections only apply for the limiting hypothetical
case of a 100\% star formation efficiency.

Gas expulsion can occur at two different times: the intra-cluster gas can
be washed out by the ionising radiation of OB stars, or later by the energy
released by supernovae explosions. As these massive stars are also expected
to eject the slowly escaping matter from which the second generation stars
form, we have two different chronologies to form a cluster with two
chemically distinct populations of low-mass stars depending on the
efficiency of the UV radiation to eject the intra-cluster gas. Indeed, in
the first case the slow winds of massive stars will accumulate in the
cluster centre where no pristine matter will be left while some dilution is
expected to occur if the ionising radiation is inefficient to expel the gas
from the cluster. This latter case would occur in clusters at the
  high-mass scale \citep{BaumgardtKroupa2008}. In both cases the second
generation stars have to be formed before the supernovae explode.

However the anticorrelation observed between Li and Na
\citep{PasquiniBonifacio2005} can help us to discriminate between these two
possibilities. Indeed some Li is present in stars with abundance anomalies
and this has to come from the intra-cluster gas itself as noted above.
Therefore gas expulsion in GC that have abundance anomalies is more likely
to be due to the supernovae than the UV radiation of massive stars. Such
models have been studied by \citet{Goodwin1997}. It should also be noted
that the slow winds of massive stars will be released at the same time as
the ionising flux from O stars, so that it would be difficult to expel the
intra-cluster gas while leaving their slow winds untouched.  We have thus
found a physical reason as to why typically more-massive GCs show more
pronounced abundance anomalies \citep[see Fig.~12 in ][]{Carretta2006} as
these clusters are more able to keep their residual gas until removal
though supernovae (see Fig.~3 in \citealp{BaumgardtKroupa2008}).

\section{Conclusions}

We have analysed the dynamical models of \citet{BaumgardtMakino2003} by
selecting two stellar populations according to their specific energies with
the aim to explain the high number of stars harbouring abundance anomalies
in GCs. The basic idea that we followed is that stars with
abundance anomalies stem from a second generation which formed centrally
concentrated out of the slow winds of rapidly rotating O stars. We find the
following results:
\begin{itemize}
\item Stars with low specific energy are initially concentrated toward the
  cluster centre and progressively diffuse into the outer parts of the
  cluster as a result of the system attempting to evolve to energy
  equipartition. About two relaxation times are needed to achieve a
  complete homogenisation between the two populations corresponding to
  several Gyr of evolution.
\item At early times, stars with high specific energy are
  preferentially lost from the cluster due to the tidal forces of the
  host galaxy. Second generation stars are only lost after several Gyr when
  they have  diffused into the outer cluster parts.
\item Any radial or dynamical differences between different 
  stellar populations would be erased over the dynamical history of old
  GCs. Even with strongly different radial distributions, as
  we assumed in this paper, GCs would today present
  homogeneous distributions. This finding could be in sharp contrast with
  other explanations for the chemical enrichment of GCs. A
  scenario which relies on the encounters between red giant stars and main
  sequence ones \citep{SudaTsujimoto2007,YamadaOkazaki2008} should produce
  more second generation stars in the cluster centre due to the high
  central density. Unfortunately, we presently do not know the true radial
  distribution of stars with abundance anomalies in GCs.
\item The typical timescale for a cluster to lose information of its
  initial conditions on the angular momentum of stars is similar to the
  time needed for radial homogenisation. Thus all initial radial and
  kinematic properties of GCs are lost after about two
  relaxation times.
\item The number ratio between the second and first stellar generation
  increases by a factor 2.5 during the cluster evolution (see also the results of \citealp{DErcoleVesperini2008}). The resulting
  ratio at old age would be too low to explain the ratio of anomalous to
  standard stars observed in GCs if we assume a standard IMF
  for the first generation stars.
\item The observed ratio would be achieved if the IMF of the first
    generation were top-heavy. This however would raise the problem of cluster
  survival \citep{BekkiNorris2006}.
\item To reproduce the observed ratio for an invariant IMF, a more powerful
  mass-loss mechanism is needed. Residual gas expulsion due to stellar
  feedback from OB stars is the prime physical candidate occurring on a
  short enough timescale and mainly removing the outer part of the cluster
  \citep{MarksKroupa2008}. Further investigations in this direction will be
  addressed in a forthcoming paper.

\end{itemize}

\begin{acknowledgements} 
  We thank an anonymous referee for comments that
  have improved this paper.  T. D. acknowledges financial support from
  swiss FNS.
\end{acknowledgements}

\bibliographystyle{aa}
\bibliography{BibADS}

\end{document}